\documentclass[prl,twocolumn]{revtex4}
\usepackage{amsmath}
\usepackage{amssymb}
\usepackage{graphicx}
\usepackage[usenames]{color}

\newcommand{\Vk}{V_{\rm eff}}
\newcommand{\ve}{V_{\rm eff}}
\newcommand{\omq}{\omega_{\rm QN}}

\definecolor{beaver}{rgb}{0.62, 0.51, 0.44}
\definecolor{applegreen}{rgb}{0.55, 0.71, 0.0}

\begin{document}

\title{Black Hole Ringing, Quasinormal Modes, and Light Rings}

\author{Gaurav Khanna} 
\affiliation{Department of Physics, University
  of Massachusetts, Dartmouth, MA 02747}

\author{Richard H.~Price} \affiliation{Department of Physics, MIT, 77 Massachusetts Ave., Cambridge, MA 02139}
\affiliation{Department of Physics, University  of Massachusetts, Dartmouth, MA 02747}

\begin{abstract}  
Modelling of gravitational waves from binary black hole inspiral has
played an important role in the recent observations of such signals. The
late-stage ringdown phase of the gravitational waveform is often associated 
with the null particle orbit (``light ring'') of the black hole spacetime. 
With simple models we show that this link between the light ring and spacetime 
ringing is based more on the history of specific models than on an actual constraining 
relationship. We also show, in particular, that a better understanding of the 
dissociation of the two may be relevant to the astrophysically interesting 
case of rotating (Kerr) black holes.
\end{abstract}

\maketitle


{\em Introduction}---Black hole ringing or ringdown (BHR) has been
important in recent gravitational wave identifications~\cite{GWPRL1,GWPRL2}.  
The late time waveform is typically identified with a quasinormal (QN) 
frequency ($\omq$). This is an association that took root in the research 
community almost a half century ago~\cite{vish,press,goebel}, and  
one that is frequently appropriate. But this is not always the case, and it 
has the great potential to be misleading. In this paper we want to 
point out the several physical and mathematical elements that are bundled 
together in the currently accepted viewpoint, to disentangle these elements, 
and to emphasize the potential for confusion and its relevance to current research.

First, QN frequencies are the complex eigenvalues of single frequency
modes; in our case, the modes of black hole perturbations. BHR
describes the damped (due to outgoing radiation) oscillation of a
black hole. ``Light rings'' (LRs) are the orbits of massless particles in
a  spacetime. These three topics often overlap, but they are fundamentally 
independent.

Current viewpoints are rooted in the history of the field. In 1970
Vishveshwara~\cite{vish} did computer simulations that revealed a
characteristic damped oscillation of quadrupole perturbations of a
Schwarzschild hole. In the early 70s, Price~\cite{price}, working with
a simple toy model for black hole mathematics, found complex poles of
the transmission function that yielded characteristic damped oscillations.  
Soon after, Press~\cite{press}, by computationally evolving perturbations, 
found damped Schwarzschild oscillations for perturbations of high multipole 
moment $\ell$.  Of particular importance to the subject's history is the 
paper by Goebel~\cite{goebel}, in which he demonstrated that Press's 
results could be understood in terms of the angular velocity and Lyapunov
exponent for the orbits of massless particles near the LR. The argument 
was heuristic, but was very appealing, and gave excellent approximations, 
even for only moderate values of $\ell$. In the early 70s this LR interpretation 
was frequently mentioned in papers~\cite{misner,smarretal} on black hole 
radiation.

Chandrasekhar and Detweiler~\cite{search} seem to have been the first to treat 
QN phenomena as unusual eigenmodes,  and to compute the Schwarzschild 
black hole's $\omq$ values as complex eigenvalues. (The fact that they can be 
complex is possible because they are eigenvalues of a non-self-adjoint 
problem~\cite{nonselfadjoint}.)

By the late 70s the idea of QNR seems to have become common in considerations of 
black hole sources of radiation~\cite{CPM1}, with LRs as part of the conceptual 
background.

One possibility of confusion lay in the fact that black hole studies
encompass two ``potentials'' with very different meaning. The analysis
of null orbits in the Schwarzschild spacetime can be understood in
terms of an ``effective potential'' for radial
motion~\cite{MTW25p6}. The peak of this potential indicates the
location of an unstable light ring, and the curvature at the peak
determines the Lyapunov exponent at that ring. By contrast, there is
also a ``curvature potential,'' that arises in the wave equation for
perturbation multipoles, such as the Zerilli
equation~\cite{zerilli}. In 1985 Schutz and Will~\cite{schutzwill},
following on related work by Ferrari and
Mashhoon~\cite{ferrarimashhoon}, demonstrated that good approximations
for values of $\omq$ could be found by applying the WKB approximation
to the peak of the curvature potential. The WKB is understood to be a
high frequency or eikonal approximation, conceptually linking this
method to the null geodesics of the spacetime, and thereby to the LRs.
In the high $\ell$ application on which Schutz and Will focused, the
curvature potential was dominated by its quasi-classical centrifugal
part. The result was the approximate equivalence of the location of
the peaks of the effective potential and the curvature potential, and
a bolstering of the apparent link between QNR and LRs. Yet other
approximations, such as the ``optical geometry''
approach~\cite{abramowicz}, have been based even more directly on the
eikonal limit.

The outcome of this multi-decade long general association of black
hole QNR and LRs is that it has been in routine use by the community
in the context of some problems (see Ref.~\cite{yangetal,cardosoetal}
and references therein), and is considered as an idea that may aid in
understanding the phenomenology of the generation of gravitational
waves in black hole binary inspiral (see Ref.~\cite{PNK} and
references therein). Moreover, it plays an important
role~\cite{eob-role} in the context of the effective-one-body (EOB)
models~\cite{eob} that are currently in use by the gravitational
observatories to generate the waveform template banks that are needed
for the signal searches.

In this paper our aim is to disentangle BHR, QNR and LRs, and to show
why such a clarification might be important to understanding the
phenomenology of gravitational waves from binary black hole inspiral.
Our clarification will include the use of
models very intentionally designed to break the connection of the
kinematic (LR) and wave (QNR) aspects of oscillations. Since it is the WKB method 
that shows the connection between these two aspects, it is to be expected that the 
WKB method fails for these models.

\bigskip
{\em Simple Models}---Let us first take on the association between the damped oscillations
of fields in a spacetime, and QNR. The disconnect between BHR and QN frequencies was 
first (to our knowledge) and most dramatically (in our opinion) demonstrated by 
Nollert~\cite{nollert} who replaced the BH mathematical problem (more specifically, the 
curvature potential) with an approximation using a set of steps. The result was a problem 
with a vastly different spectrum of QN frequencies, but almost identical BHR. Other such models 
have also been presented in the recent research literature~\cite{cardosoPRL,konoplyazhidenko}.

Turning to the more tangled issue of QNR and LRs, we shall exploit the convenience of 
gravity-free spherically symmetric wormholes as simple examples. The metric for such a 
spacetime is
\begin{equation}
  ds^2=-dt^2+dx^2+r^2(x)\left(d\theta^2+\sin^2\theta\,d\phi^2\right)\,.
\end{equation}
All of the properties of this wormhole spacetime are contained in 
the function $r(x)$. 

It is straightforward to show that the function $\ve\equiv1/r^2(x)$
serves as an effective potential in the same sense as the effective
potential for the Schwarzschild spacetime~\cite{MTW25p6}. In particular
a circular null orbit requires that $dr/dx=0$, and  such an orbit
is unstable if $d^2r/dx^2>0$. For our first example, we choose our
$r(x)$, and thereby choose the effective potential $\ve$, to be given by
\begin{widetext}
\begin{equation}\label{eq:Fdef}
\Vk(x)\equiv\frac{1}{r^2(x)}=\frac{\pi
}{\sqrt{2}}-\left(-\frac{1}{\sqrt{2}}+\frac{k}{2}\right)
\text{Tan}^{-1}\left(1-\sqrt{2}
  x^2\right)-\left(\frac{1}{\sqrt{2}}+\frac{k}{2}\right)
\text{Tan}^{-1}\left(1+\sqrt{2} x^2\right)\,.
\end{equation}
\end{widetext}
This function is symmetric in $x$ and has the property that
$r\rightarrow x+{\cal O}(1/x)$ as $x\rightarrow\pm\infty$, thus the
spacetime is asymptotically flat.  This wormhole has a minimum radius
at $x=0$, where there is an unstable LR. If $k<2$, this is the only
LR, but for $k>2$ there are two other unstable LRs located symmetrically 
around the central LR.  

The effective potential determines the curvature potential for a given
multipole.  For a scalar mutipole perturbation,
$\Psi=r^{-1}\psi(x,t)Y_{\ell,m}(\theta,\phi)$, the sourceless wave
equation $g^{\mu\nu}\Psi_{,\mu;\nu}=0$ takes the form
\begin{equation}
\frac{\partial^2\psi}{\partial t^2}-\frac{\partial^2\psi}{\partial x^2}
+V_{\rm curv}(x)\psi=0\,,
\end{equation}
where the curvature potential $V_{\rm curv}$ turns out to be
\begin{equation}\label{eq:Vcurv}
  V_{\rm curv}\!\!=\ell(\ell+1)\Vk(x)\!\!- \frac{1}{2\Vk}\left[\frac{d^2\Vk}{dx^2} 
-\frac{3}{2\Vk}\!\left(\frac{d\Vk}{dx}\right)^2
\right]\,.
\end{equation}

\begin{center}
\begin{figure}[h]
  \includegraphics[width=.35\textwidth ]{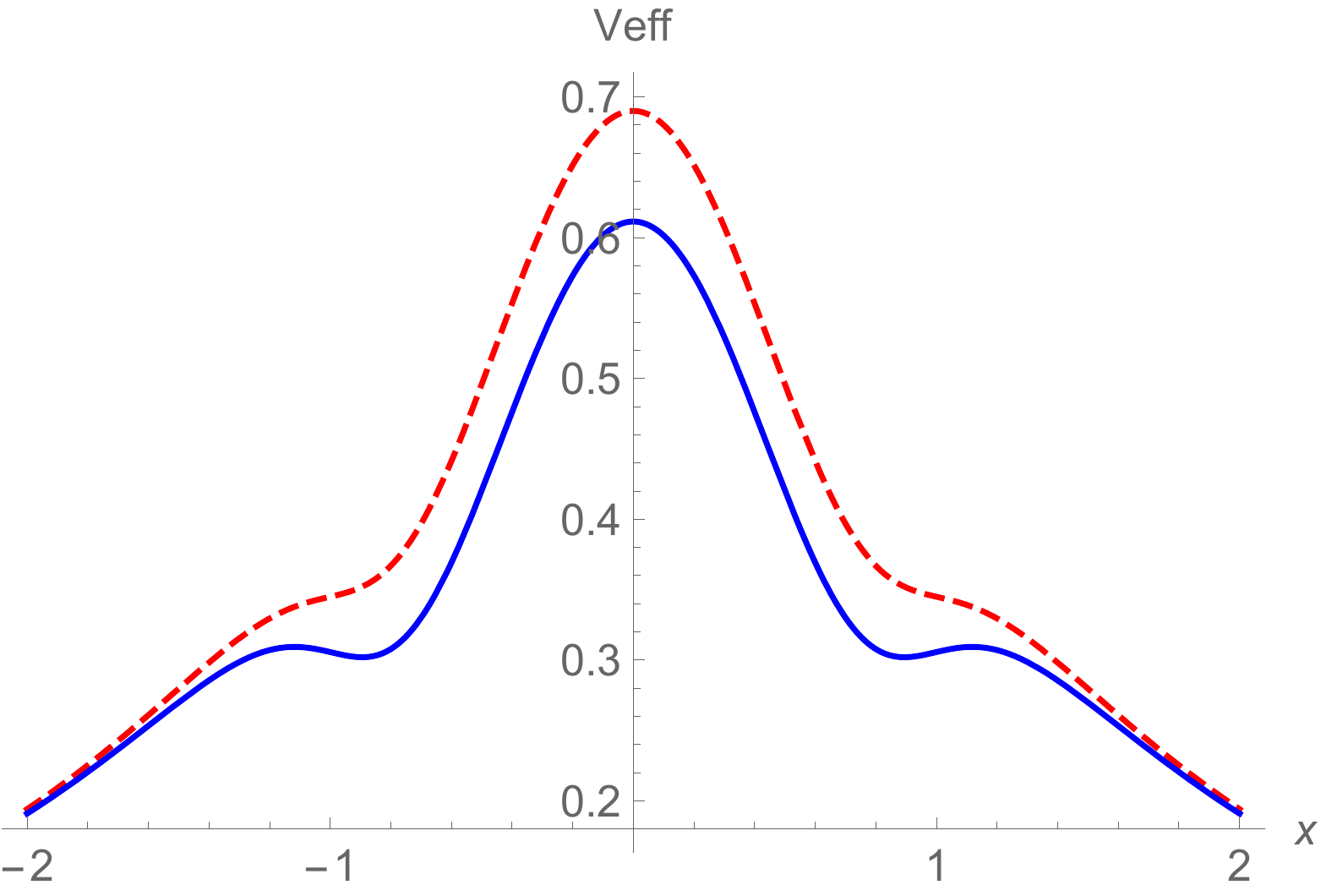}\\
  \includegraphics[width=.35\textwidth ]{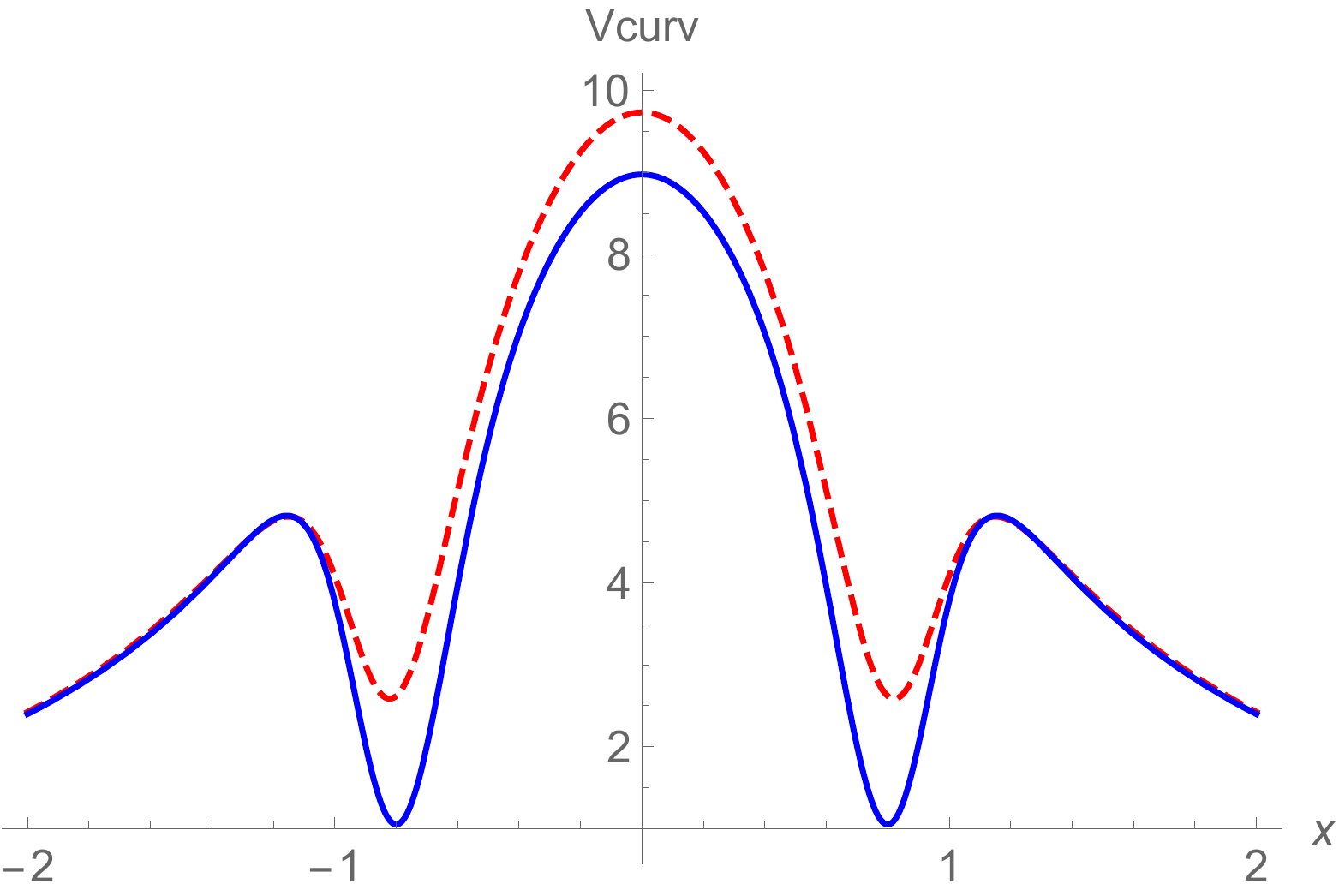}
  \caption{The effective  and $\ell=3$ curvature  potentials for models 
with $k=1.95$ (dashed) and $k=2.05$ (solid).}
  \label{fig:2Vs}
  \end{figure}
  \end{center}

A particular example is shown in Fig.~\ref{fig:2Vs}: the effective and
curvature potentials for $k=1.95$ and $2.05$.  The top graph shows
$\Vk$ for each of the $k$ values. Of particular importance is the fact
that for the smaller $k$ value there is only the $x=0$ central LR; for
the larger $k$ there are additional unstable LRs at $x\approx\pm1.12$. 
The lower graph  shows the $\ell=3$ curvature potentials; these potentials 
determine the values of $\omq$. The plot shows that the curvature potentials 
are qualitatively the same, even quantitatively similar, independent of the 
number of LRs.

\begin{center}
\begin{figure}[h]
  \includegraphics[width=.45\textwidth ]{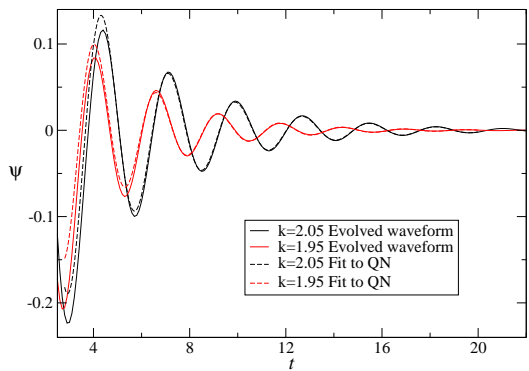}
  \caption{Waveforms for the Eq.~(\ref{eq:Fdef}) model, with 
$k$ parameters $1.95$ and $2.05$. The solid curves show the computed evolution
of initial data, while the dashed curves show the fit to the least damped QN 
oscillation of each model.}
  \label{fig:comparo}
  \end{figure}
  \end{center}

Since the curvature potential determines the oscillations of the spacetime, we
would expect that the two models in Fig.~\ref{fig:2Vs} will have very similar
oscillations. Figure~\ref{fig:comparo} shows that this expectation is met. 
The oscillations shown, furthermore, fit very well to waveforms with the QN
values $\omq=2.440+i\,0.3203$ (for $k=1.95$) and $\omq=2.256+i\,0.2526$ (for $k=2.05$)
found by an eigenvalue search in the complex plane adopted from that of 
Ref.~~\cite{search}. The waveforms are clearly qualitatively similar, and 
quantitatively not very different. They are, furthermore, clearly QNR.

It is of some interest  to apply the Schutz-Will~\cite{schutzwill} WKB approximation 
to the models we consider using the formula~\cite{schutzwill}
\begin{equation}
  \omega^2_{\rm QN}=
V_{\rm curv}-i(n+\textstyle{\frac{1}{2}})\sqrt{-2{d^2V_{\rm curv}}/{dx^2}\;}\,,
\end{equation}
where it is understood that $V_{\rm curv}$ is to be evaluated at the
peak. In doing this it is important to keep in mind that the 
limitation of that approximation is that, in an equation with the form
\begin{equation}
\frac{\partial^2\psi}{\partial x^2}+\left(\omega^2
+V_{\rm curv}(x)\right)\psi=0\,,
\end{equation}
the relationship 
\begin{equation}\label{WKBcond}
  \left|\frac{dV_{\rm curv}(x)}{dx}\right|\ll \left[\omega^2+
V_{\rm curv}(x)
\right]^{3/2}
\end{equation}
holds throughout the region to which the approximation is applied~\cite{MW}.
We can simplify this slightly by using the fact that the value of 
the WKB frequency will always be roughly equal to the peak of the curvature
potential and this we can approximate a lower bound on the right side of 
Eq.~\eqref{WKBcond} by $|\omega|^3$. 

In applying this criterion to the models illustrated in
Fig.~\ref{fig:2Vs} we can see immediately that the worst violation
occurs around $x\approx\pm0.63$ where $\left|{dV_{\rm
    curv}(x)}/{dx}\right|$ is 20.3 and 16.2, for the $k=2.05$ and
$k=1.95$ cases, respectively. The values of $|\omega|^3$, on the other
hand, are 11.5 and 14.5 for the $k=2.05$ and $k=1.95$ cases,
respectively, so the condition for validity of the WKB approximation
is badly violated. The actual values given by the Schutz-Will approximation
is in error by around 50\%, which is smaller than what one might have guessed.

\begin{figure}[h]
  \begin{center}
  \includegraphics[width=.35\textwidth ]{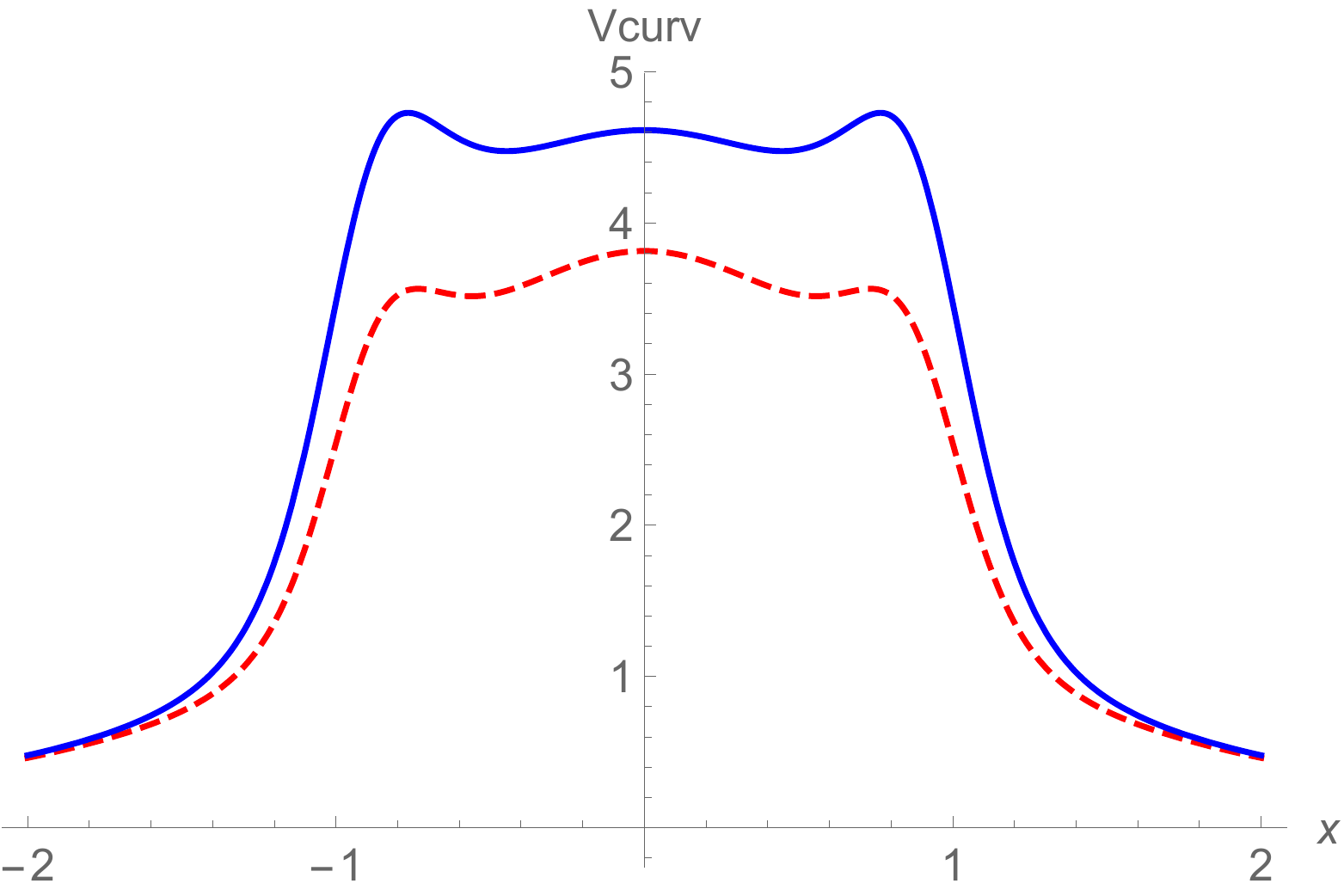}
  \caption{The curvature potentials for our wormhole models with  
  $k=0.8$ (dashed) and $k=0.2$ (solid).}
  \label{fig:Batmen}
  \end{center}
  \end{figure}

For another example, we choose a scalar dipole ($\ell=1$), 
and $k$ values 0.2 and 0.8. The curvature potentials are shown 
in Fig.~\ref{fig:Batmen}.
The location of the larger peak should be noted. For $k=0.8$ the 
larger peak is at $x=0$; for $k=0.2$, they are at $x\approx\pm1$.

\begin{figure}[h]
  \begin{center}
  \includegraphics[width=.45\textwidth ]{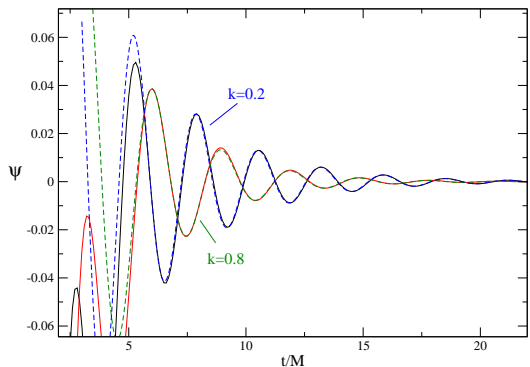}
  \caption{The ringdown phase of the time-domain scalar field waveforms 
   in our example wormhole spacetimes for $k=0.8$ and $k=0.2$, matched to our 
   precision QNR results from the frequency plane eigenvalue search.}
  \label{fig:evolk0p20p8}
  \end{center}
  \end{figure}

With a search in the frequency plane using the method of
Chandrasekhar and Detweiler~\cite{search} we have been able to achieve
considerable precision in determining the $\omq$ for these two cases.
Figure~\ref{fig:evolk0p20p8} shows the evolution of initial data in
each of the backgrounds, and shows that the fit of the late-time evolved
waveforms to these $\omq$ is excellent. There is, then, no
question that the $\omq$ are meaningful, and that they describe
the waveform that develops in these spacetimes.  Table~\ref{table:QNs}
compares these $\omq$ to values given by WKB and LR
approaches. The second column gives true least damped $\omq$,
the frequencies found by the above mentioned frequency domain eigenvalue 
search, which are also the frequencies that fit the evolved waveform.
The third column and fourth columns give WKB results. 
The third column gives the result of applying this for the peak 
at $x=0$; the fourth column gives the result for the peaks at $|x|\approx1$.  
Lastly, the fifth column is the prediction of the $\omq$ based on the 
LR analysis given by Cardoso {\it et al.}~\cite{cardosoetal} 
(see, in particular, Eqs.~(2) and (40) of that reference). It is worth 
emphasizing again, that this LR analysis is based on the kinematic potential,  
{\em not} the curvature potential.  We have applied that LR analysis at 
$x=0$, the location of the only LR in these models. (Recall that the peaks 
in $V_{\rm curv}$ at $x\approx\pm1$ are not related to LRs; for the functions 
$r(x)$ given by Eq.~(\ref{eq:Fdef}) models with $k<2$ have only the central 
LR.)

\begin{widetext}
\begin{center}
\begin{table}[h]
   \begin{tabular}{|c|c|c|c|c|}
\hline
$k$ &
Actual&
WKB at $x=0$ &  
WKB at $x\approx\pm1$ &  
LR\\ \hline
0.2&
$2.3638+i\,0.2901$&
$2.1646+i\,0.2686$&
$2.3243+i\,0.8214$&
1.4368+$i\,0.3480$\\
0.8&
2.1364+$i\,0.3633$&
1.9841+$i\,0.3501$&
1.9966+$i\,0.6502$&
1.2622+$i\,0.3962$\\ \hline
\end{tabular}
 \caption{
 Computed QN frequencies from different approaches, as described in 
 the text, for the $k = 0.2, 0.8$ wormhole model cases. It is clear 
 that while the WKB approximation performs reasonably well (especially 
 for the real part) and LR approach performs very poorly. }\label{table:QNs}
\end{table}
\end{center}
\end{widetext}

The results in this table are quite telling. The WKB approximation applied at 
the $x=0$ peak gives reasonable approximations for the true values. The WKB
approximation applied to the  peaks gives more or less equally good
approximations for the real part of the $\omq$, but due to the narrowness
of the peaks, gives values of the damping that are significantly too high.
It is interesting that for the $k=0.2$ model, the peaks at $x\approx\pm1$
are higher than at $x=0$, and yet the WKB approximation at $x=0$ is better.
An intuitive explanation for this is that the wavelength of the relevant 
mode is much wider than the entire potential; thus, the results are likely 
to be relatively insensitive to the small scale peaks. 

What is most important to notice is that the approximation based on
the LR is far from correct. This is not surprising in retrospect,
since the kinematic potential  given by
Eq.~(\ref{eq:Fdef}) 
is dramatically different from the curvature potential in
Eq.~(\ref{eq:Vcurv}), shown in Fig.~\ref{fig:Batmen}. Much of the 
published literature in this context is in the ``geometric optics''  
i.e., large $\ell$, limit wherein the curvature potential is dominated by 
the centrifugal term. We have considered here curvature potentials that 
are not the single-peak potentials for which the LR argument works well. 
And it is the curvature potential that governs waves, and hence determines 
everything about the QN phenomena for a spacetime.

\begin{figure}[h]
  \begin{center}
  \includegraphics[width=.35\textwidth ]{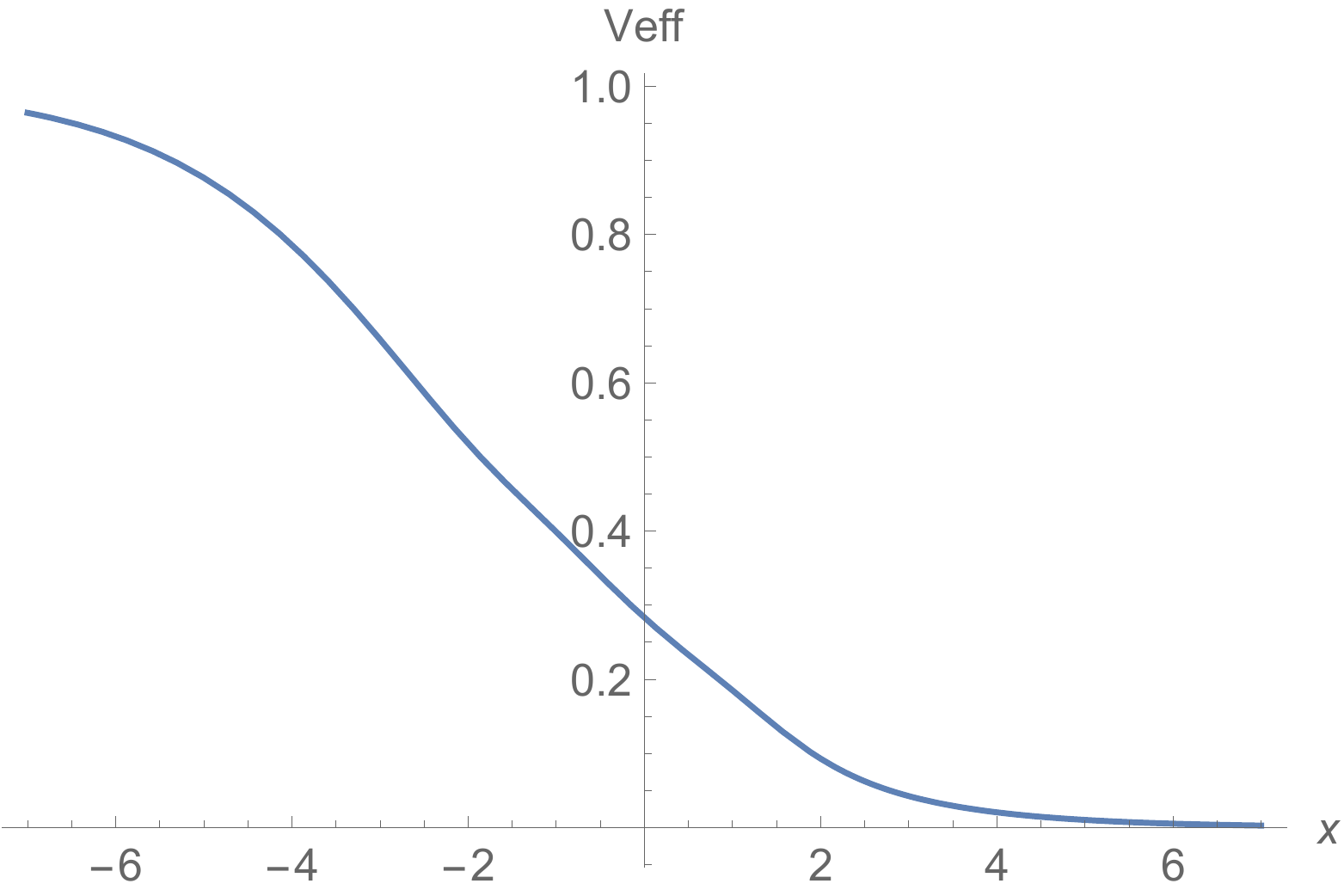}\\
  \includegraphics[width=.35\textwidth ]{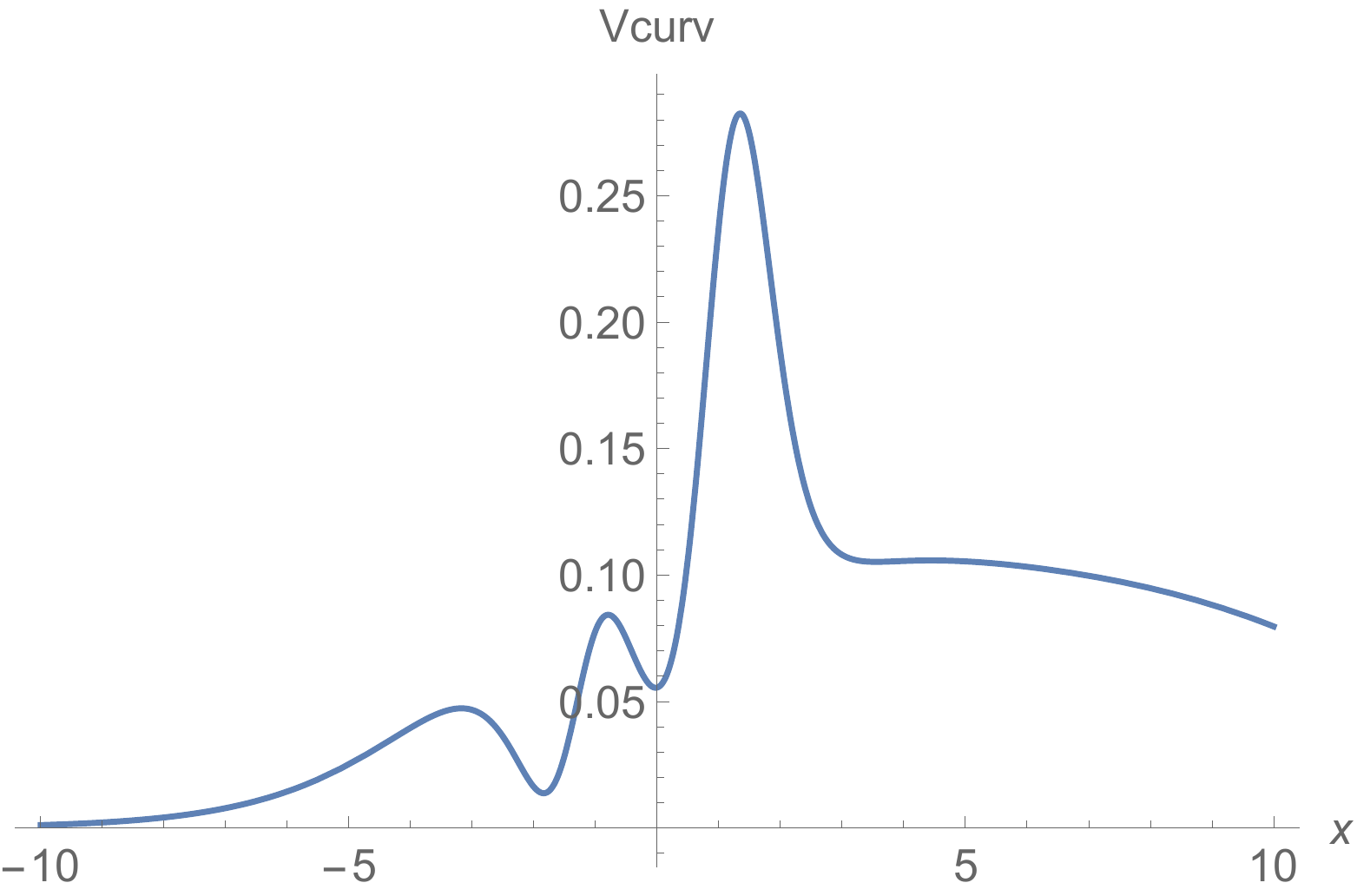}
  \caption{ The effective potential (top), and $\ell=0$ curvature
    potential (bottom) for scalar waves in a spacetime defined by
    Eq.~(\ref{eq:NoLRF}). This model has no LR, but exhibits QNR. }
  \label{fig:NoLRF}
  \end{center}
  \end{figure}

\begin{figure}[h]
  \begin{center}
  \includegraphics[width=.45\textwidth ]{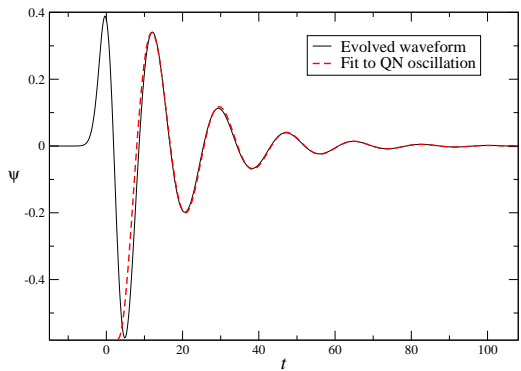}
  \caption{ 
Evolution of initial data for a wormhole model with no LR.}
  \label{fig:wave2o3}
  \end{center}
  \end{figure}

The previous examples help to weaken the link between LRs and oscillations 
in curved spacetimes. Next, we weaken that even further with a model defined by
\begin{equation}\label{eq:NoLRF}
\Vk=\frac{1}{r^2(x)}=\frac{e^{-2 x/3}}{4+e^{-2 x/3}}
+\frac{1}{12+x^2+x^4}\,,
\end{equation}
and pictured in Fig.~\ref{fig:NoLRF}. As the figure shows, this
wormhole approaches a cylinder (constant radius $r=1$) asymptotically,
as $x\rightarrow-\infty$. What is most important, and immediately
apparent in the top plot in Fig.~\ref{fig:NoLRF}, is that there is no LR, 
since there is no point at which $dr/dx=0$. Despite this, the associated 
curvature potential for monopole waves, shown in Fig.~\ref{fig:NoLRF} (bottom) 
has the peaked form that suggests that the spacetime will have QNR.
Figure~\ref{fig:wave2o3} shows that this is indeed the case. The solid
curve shows the result of the computer evolution of scalar initial data. 
That evolved waveform is compared to the fit to a QN oscillation with
$\omq=0.356+i\,0.060$, the value found with a search in the complex plane 
for the least damped monopolar QN mode. The figure leaves no doubt that 
this spacetime, despite the absence of {\it any} LR, exhibits QN 
ringing~\cite{asympflat}. 

For this ``no LR'' model, we can apply the WKB approximation to the
major peak of the curvature potential at $x\approx1.35$. We expect the
WKB model to fail, and it is no surprise that it fails badly, giving the 
0.692731+i\,0.4442 which is off by a considerable factor from the correct value.

\begin{figure}[h]
  \begin{center}
  \includegraphics[width=.45\textwidth ]{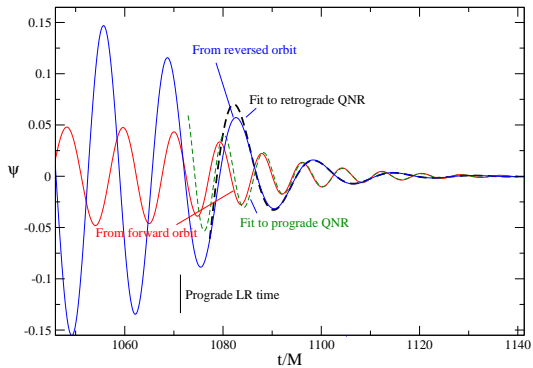}
  \caption{ 
  Ringdown stage scalar field waveforms from a test particle with scalar 
  charge falling into a Kerr black hole with $a/M = 0.9$ on a prograde 
  geodesic, and also on a retrograde ``reversed'' non-geodesic path. It is 
  clear that the prograde case excites the $+m$ QNR, while in the retrograde 
  case the $-m$ mode is excited.  }
  \label{fig:scalarKerr}
  \end{center}
  \end{figure}

\medskip
{\em Possible relevance to Kerr ringdown}---Our goal in this section
is to show that the unreliable LR/QNR association may mislead research
in the phenomenology of binary inspiral. As a plausible example of the 
failure of this association we present, in Fig.~\ref{fig:scalarKerr}, the
result of a particle perturbation evolution code~\cite{teukcode}
representing a scalar-charged particle spiraling into a Kerr black
hole with $a/M=0.9$. Two curves are presented. One shows the scalar
radiation generated by a prograde geodesic (``forward'') equatorial
orbit for per-particle-mass energy $E=0.84$, and angular momentum
$L/M=2.1$. The second curve is the radiation from the (non-geodesic)
``reversed'' orbit resulting from the reversal of the angular direction of 
the ``forward'' orbit. Both orbits start with the particle located 
at $r=2.5\,M$. The ``junk radiation'' attending the birth of the particle 
quickly dissipates, and is irrelevant to our considerations here. 
From the published results of Berti {\it et al.}~~\cite{bertietal} we 
have that for such a black hole, the least damped quadrupole scalar 
$\omq$ are $(0.78164 + i\,0.06929)/M$, for $m=+2$, and 
$(0.38780 + i\,0.09379)/M$ for $m=-2$. The figure shows the excellent 
fit of the late-time forward/reverse radiation to the $m=+2/-2$ modes 
respectively.

It is intuitively appealing that the retrograde, reversed orbit
generates $m=-2$ QNR, and it is tempting to associate this with the
retrograde LR. But there is an important barrier to this association.
QNR is a phenomenon of the curvature potential, and the excitation
of QNR by an infalling particle can be traced to the passage of the
particle past a feature of the curvature potential~\cite{PNK}. In the common
viewpoint this is equivalent to passing the LR (see Ref.~\cite{PNK} and 
references therein). This viewpoint does not appear to apply to the example in 
Fig.~\ref{fig:scalarKerr}. For $a/M=0.9$, the retrograde LR is at $r=3.91027\,M$, 
and our particle source, whether in forward or reverse orbit, starts at $r=2.5\,M$ 
and hence spends no time at the retrograde LR, the location that one might 
associate with the retrograde QNR. Note that the prograde LR is at $1.55785\,M$.\  
The short vertical bar in Fig.~\ref{fig:scalarKerr} shows the time,
$\sim1070\,M$ at which the particle reaches this prograde LR. One can
interpret this to mean that the radiation from the forward orbit is
associated with the prograde LR, but how does one explain the fact
that the reverse-orbit radiation starts at around the same time?

One plausible answer is simply that the QNR/LR association must be treated 
with caution and even skepticism.

{\em Conclusions}---In many cases the oscillations of a black hole, or
other spacetime, is a manifestation of QNR, a complex eigenmode for radiation 
in the spacetime. In the development of black hole perturbation studies quasinormal 
ringing has been successfully linked to the orbits of null particles, the LRs. 
We have shown that this is a weak link by presenting models in which the QNR 
is clearly not linked to such LRs, even a model in which there is very clear 
QNR in a spacetime with no LR.

We also noted that this dissociation of QNR and LRs may explain some of 
the phenomenology of binary inspiral radiation, especially in connection with the 
QN mode that has been thought to be related to the retrograde light ring.

\medskip
{\em Acknowledgments}---We would like to acknowledge many discussions 
with Alessandra Buonanno and Scott Hughes on this topic over the years, 
and also for their feedback on an earlier version of this paper.   
G.K. acknowledges research support from NSF Grants No. PHY-1414440 and
No. PHY–1606333, and from the U.S. Air Force agreement No. 10-RI-CRADA-09.

\medskip

\begin{thebibliography}{11}
\bibitem{eob-role} In the EOB approach, the early (inspiral-plunge) signal 
is matched to a late (merger-ringdown) superposition of QN modes at the LR.
\bibitem{asympflat} The model in Eq.~(\ref{eq:NoLRF}) is less than
  ideal in that it does not obey the asymptotically flat condition
  $r\rightarrow x$ as $x\rightarrow\infty$. We have alternate models
  that are asymptotically flat, but none shows as well as
  Fig.~(\ref{fig:wave2o3}) a high ratio of QN oscillations to damping,
  and hence the opportunity to demonstrate the QN fit.
\end{thebibliography}
\end{document}